\documentclass[twocolumn,prb,nofootinbib]{revtex4}
\usepackage{graphicx}
\usepackage{epsfig}
\usepackage{dcolumn}
\usepackage{bm}
\usepackage{amsmath}
\usepackage{amssymb}

\newcommand{\ket}[1]{\ensuremath{\,|{#1}\rangle}}

\begin{document}


\title{Donor-donor interaction mediated by cavity-photons \\
and its relation to interactions mediated by excitons and polaritons}

\author{G. F. Quinteiro}

\affiliation{Departamento de F\'{\i}sica, FCEN, Universidad de
Buenos Aires,Ciudad Universitaria, Pabell\'{o}n I C1428EHA Buenos
Aires, Rep\'{u}blica Argentina
\\ Instituto de Ciencias, Universidad Nacional de General Sarmiento,
J. M. Gutierrez 1150 C.P. 1613 Los Polvorines, Pcia. de Buenos
Aires, Rep\'{u}blica Argentina}

\date{\today}

\begin{abstract}
I report theoretical predictions of two models of donor-donor
indirect interaction mediated by photons in zero- and
two-dimensional cavities. These results are compared to previously
studied cases of indirect interactions mediated by excitons and/or
polaritons in bulk semiconductor and two-dimensional cavities. I
find that photons mediate an Ising-like interaction between donors
in the same manner polaritons do, in contrast to the Heisenberg-like
interaction mediated by exciton. For the particular case of a
two-dimensional cavity, the model shows that the dependence on
distance of the donor-donor coupling constant is the same for
photons and polaritons when the donor-donor distance is large. Then,
it becomes clear that photons are responsible for the long range
behavior of the polariton indirect interaction.
\end{abstract}

\maketitle


\section{Introduction}
\label{Sec_intro}

The advances in the fabrication of low dimensional structures have
opened new areas of research in condensed matter physics. These new
man-made systems pose challenges in terms of basic science, but also
permit further technological developments. The interest and current
progress in the fabrication and characterization of zero- (0D) and
two-dimensional (2D) cavities containing quantum dots (QD) and
impurities is evidenced from the work of several experimental groups
around the world.~\cite{Ralthmaler_04}

Indirect interactions (II) in solid state physics are a well-known
phenomenon. The first investigations date back to the 1950s with the
study of the II between nuclear spins mediated by conduction
electrons in metals~\cite{Ruderman} and in
insulators~\cite{Bloembergen}, process called RKKY. Closer to
present days we find the Optical RKKY~\cite{piermarocchi_02,
piermarocchi_04}, where the II is mediated by optically excited
excitons in semiconductors. As already mentioned, the developments
in nanoscience call for the revision and extension of previous
works. Based on the ideas mentioned above and other related studies,
investigations in the II in low dimensional structures are being
carried out to answer basic questions as well as to suggest
applications. Examples of these works are the II in QDs embedded in
microdisc structures as a scheme for quantum
computing~\cite{Imamoglu99}, the ferromagnetic ordering induced by
the II~\cite{JFR04}, the II between impurities in a QD inside a $0$D
cavity~\cite{Chiappe_05}, the II between impurities close to QDs
embedded in quantum wells as a scheme for quantum
operations~\cite{quinteiro_05}, and the II in $2$D cavities mediated
by polaritons~\cite{quinteiro_06}. As can be seen, a major field of
application of these concepts is quantum computing, since the
optical control of single impurities or quantum dots promises to be
a fast and suitable tool for implementing quantum operations.

This article presents theoretical predictions for two closely
related systems exhibiting donor-donor II mediated by photons. The
systems are a $0$D and a $2$D cavities with embedded QDs each
coupled to a donor. To the best of the author's knowledge, no
previous works report on either the Optical RKKY-like interaction
mediated by virtual photons in $2$D cavities with QDs, or the exact
solution of the II mediated by photons in $0$D cavities. In
addition, the results are compared to previously studied cases of
indirect interactions mediated by excitons and/or polaritons in bulk
semiconductor and $2$D cavities. The article is organized as
follows: First, a common framework for both models is given in
Section~\ref{Sec_gencon}. Section~\ref{Sec_2D} and \ref{Sec_0D}
describe, respectively, the $2$D and $0$D models together with their
results. Section~\ref{Sec_Conclusions} presents the general
conclusions of this work.

\section{General considerations}
\label{Sec_gencon}

Here I summarize the main features common to both theoretical
models. The systems consist of a cavity, either $0$D or $2$D,
containing two QDs each one coupled to a donor
impurity.~\footnote{The impurity may be placed inside the QD or
alternatively, it may be located close to the QD by
modulation-doping so as to ensure direct interaction between it and
the exciton in the QD.~\cite{Siegert05, Besombes04}} The neutral QDs
may be excited by a cavity photon, and the resulting exciton may
interact with the impurity via an exchange interaction between the
electron in the donor and the electron in the exciton -- for donors,
the hole-electron exchange is typically smaller than the
electron-electron exchange and is thus neglected. Due to quantum
confinement, the heavy hole and light hole levels split; then, one
can restrict the study to the heavy hole level only which spans a
four dimensional space for the exciton states in the QD: two
optically active and two dark states.

I first treat the case of a $2$D cavity, and solve the problem
analytically by perturbation theory. Then, the $0$D cavity model is
studied by an exact numerical diagonalization of its Hamiltonian.

\section{The donor-donor interaction mediated by virtual photons in a $2$D cavity}
\label{Sec_2D}

\begin{figure}[h]
  \centerline{\includegraphics[scale=0.5]{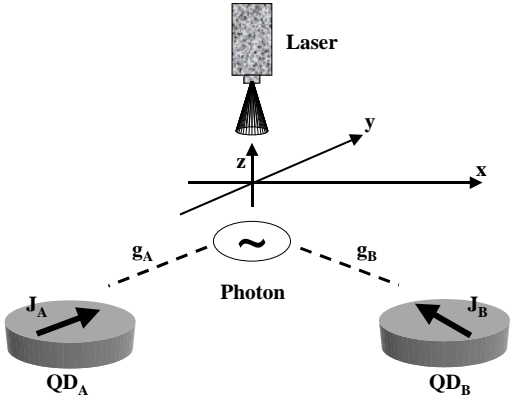}}
  \caption{The system consists of a 2D-cavity (not drawn)
  and two QDs each coupled to a donor impurity. The cavity is
  excited from outside by an off-resonance
  monochromatic electro-magnetic field propagating normal to the
  cavity in the $z$-direction. \label{fig_2D}}
\end{figure}

The model consists of a $2$D cavity with two embedded QDs, each of
them coupled to a donor. The system is optically excited by an
off-resonance monochromatic laser field propagating parallel to the
$z$-axis (normal to the cavity), which excites virtual cavity
photons (Fig~\ref{fig_2D}).

Inside the cavity, the quantized electric field with wave-vector
\mbox{$\textbf{m}=\textbf{k}+\textbf{q}$} of canonical spherical
angles $\{\phi,\theta\}$ having in-plane/$z$ component
$\textbf{k}/\textbf{q}$ is
\begin{eqnarray}
\label{Eq_Field}
    \textbf{E} &=& i \sum_{\chi k q} \left( \frac{\hbar\,\omega_k}
    {2\,\epsilon\, V} \right)^{1/2}
    \textbf{n}_{\chi k\,q}\, c_{\chi k\,q}\,
    e^{-i(\omega_k t-{\bf k}{\bf R})} \nonumber \\
&&  + h.c. \,,
\end{eqnarray}
where $\textbf{n}_{\chi k\,q} = R_{zy} (\phi,\theta)
\,\boldsymbol{\varepsilon}_\chi$ is the circular polarization vector
resulting from a rotation of the vector
\mbox{$\boldsymbol{\varepsilon}_\pm = \hat{x}\pm i\hat{y}$},
$c_{\chi k\,q}$ is the annihilation operator for cavity photons with
frequency $\omega_k = \frac{c}{n} \sqrt{k^2+q^2}$, and $V$ is the
cavity volume. Due to the spatial confinement in the $z$ direction,
the corresponding momentum is restricted to two possible values
${\bf q}=\pm q \,\hat{z}$ -- inversely proportional to the width of
the cavity -- while the in-plane momentum $\bf k$ has a
quasi-continuum spectrum. The phase factor $e^{iqz}$ has been
eliminated by taking the quantum well at $z=0$.

\subsection{The Hamiltonian}

The complete Hamiltonian involves the bare energy ($H_0$) and the
interactions: laser - cavity photon ($H_{LC}$), cavity photon - QD
exciton ($H_{CX}$), and QD exciton - donor ($H_{XS}$).\\

\textbf{The bare Hamiltonian:}
\begin{equation}\label{Eq_H0}
    H_0=E_{X}^A + E_{X}^B + \hbar \sum_{\chi\,k\,q}
    \omega_k\, c_{\chi k\,q}^\dag c_{\chi k\,q}~,
\end{equation}
with exciton energy $E_{X}^i < \hbar \omega_k$, for the $i=A/B$ QD;
the ground state energy of the donors is taken equal to zero.\\

\textbf{The laser - cavity photon Hamiltonian: } In accordance with
the quasimode approach, useful for high-Q cavities~\cite{Quasimode},
the interaction between the laser and the cavity photon is
represented by a semi-classical field that creates/annihilates
cavity photons with momentum in $z$ equal to $\pm q$
\begin{equation}\label{Eq_HLC}
    H_{LC}=\hbar \sqrt{{\cal A}} \sum_{\sigma q} {\cal{V}}_\sigma
    e^{i \omega_L t}\, c_{\sigma 0 q} + h.c.~,
\end{equation}
where ${\cal A}$ is the area of the $2$D cavity, and
${\cal{V}}_\sigma$ the coupling constant.  \\

\textbf{The cavity photon - QD exciton Hamiltonian:} The Schr\"{o}dinger
representation of the interaction between the cavity mode Eq.~\ref{Eq_Field}
and the exciton taken in the dipole moment approximation is
\begin{eqnarray}
\label{Eq_HCX}
    H_{CX} &=&
        i\,\frac{g}{\sqrt{{\cal A}}}\, \sum_{\chi \sigma k q p}
        (\textbf{n}_{\chi k q}\cdot
        \,\textbf{d}_\sigma^*) \, m^{1/2}
        e^{i\textbf{k}\textbf{R}_p}
        ~b^{(p)\dag}_{\sigma} c_{\chi k q} \nonumber \\
    &~& + h.c.\,,
\end{eqnarray}
with $g$ the coupling constant, and $m=|{\bf m}|$.
$b^{(p)\,\dag}_\sigma$ is the creation operator for excitons in the
$p=A/B$ QD with the condition: \mbox{$\sigma+ \equiv (e\downarrow,
h\uparrow$)} and \mbox{$\sigma- \equiv (e\uparrow, h\downarrow$)}
for the spins of the electron (e) and the hole (h). Assuming the
quantization axis for the excitons in $z$, the dipole moment is:
$\textbf{d}=\mp \frac{1}{\sqrt{2}}\,(\hat{x} \pm i \hat{y})$. The
scalar product is
\begin{eqnarray*}
        \textbf{n}_{\pm k q} \cdot \textbf{d}_\pm^*
    &=& \frac{1}{2}\,e^{\pm i\phi}\left(1 + \frac{{\bf q}\cdot \hat{z}}{m}\right) \\
        \textbf{n}_{\pm k q} \cdot \textbf{d}_\mp^*
    &=& \frac{1}{2}\,e^{\mp i\phi}\left(1 - \frac{{\bf q}\cdot \hat{z}}{m}\right) \,,
\end{eqnarray*}
Notice that the interaction is valid for all angles $\{\theta, \phi\}$.\\

\textbf{The QD exciton - donor Hamiltonian:}
\begin{equation}\label{Eq_XS}
    H_{XS}= \sum_{p \, \alpha \alpha^\prime \beta} J^{(p)}\,
    s_{\alpha\alpha^{\prime}}\cdot\hat{s}^{(p)} ~~
    \hat{b}^{^(p)\dag}_{\alpha \beta}\hat{b}^{(p)}_{\alpha^\prime \beta} + h.c.~,
\end{equation}
where $\hat{b}^{^(p)\dag}_{\alpha \beta}$ has electron spin $\alpha$ and hole
spin $\beta$ (same operator as in Eq.~\ref{Eq_HCX}, but with the spin
index explicitly written for the electron and the hole), and the tensor
\begin{displaymath}
\mathbf{[s_{\alpha\alpha^{\prime}}]} = \frac{\hbar}{2} \left(
\begin{array}{cc}
\hat{z} & \hat{x}-i\hat{y} \\ \hat{x}+i\hat{y} & -\hat{z}
\end{array} \right)\, ,
\end{displaymath}
with, for example, $s_{12}=s_{\uparrow\downarrow}$.\cite{quinteiro_06}
Notice that the interaction does not change the spin of the hole.

\subsection{The effective donor-donor Hamiltonian}

I seek to derive a Hamiltonian expression for the donor-donor
interaction that contains operators only for the spin degree of
freedom of each impurity. The procedure involves moving to a
rotating frame to eliminate the time dependence from $H_{LC}$,
applying second order perturbation theory to $H_{LC}$, and finding
the correction to the unperturbed Hamiltonian $H_0$, keeping the
lowest order in $H_{XS}$ and $H_{XC}$ (See Appendix~\ref{App_Heff}
for details). It is worth mentioning that the off-resonance
excitation of the system makes possible the use of perturbation
theory, since the detuning --which is a controllable parameter--
enters the expansion coefficient.

\begin{figure}[h]
  \centerline{\includegraphics[scale=0.5]{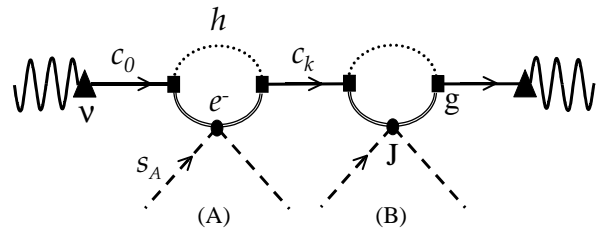}}
  \caption{Diagrammatic representation of the lowest order contribution
  (${\cal{O}}[(G^{0\,2}_X\,g^2\,J\,F)^2]$) to the indirect interaction
  between two donors. Loops are for excitons
  in each (A/B) QD, solid lines are for cavity photons, wavy lines
  are for the laser field, and dash lines are for donors. The coupling
  constants $\{\nu, J, g\}$ are represented by triangle, circle and
  square vertices, respectively.
  \label{fig_2Ddiagram}}
\end{figure}

The first relevant term that gives rise to a process that correlates
both spins is (the same process exchanging $A\leftrightarrow B$ must
also be considered): $1st)$ The extra-cavity EM-field shines on the
cavity coherently exciting a virtual cavity photon, $2nd)$ the
cavity photon creates an exciton in QD $A$, $3rd)$ the exciton
interacts with the donor spin A, $4th)$ the exciton in QD $A$
annihilates and a cavity photon is created, $5th)$ the cavity photon
creates an exciton in QD $B$, $6th)$ the exciton interacts with the
donor spin B, $7th)$ the exciton in QD $B$ annihilates and creates a
cavity photon, $8th)$ the cavity photon is deexcited; see
Fig.~\ref{fig_2Ddiagram}. It should be kept in mind that the whole
process is driven by an off-resonance excitation, and thus no energy
absorption occurs. The final expression for the effective
Hamiltonian corresponding to this process is
\begin{widetext}
\begin{eqnarray}
\label{Eq_Heff}
H_{eff} &=& \left\{ \frac{2\,\hbar^3\,q\,n}{c}
    \,\frac{|{\cal{V}}_{\sigma+}|^2 +|{\cal{V}}_{\sigma-}|^2}
    {\delta_C^2}
    \,\frac{J^{(A)}g^{(A)\,2}\,J^{(B)}g^{(B)\,2}}
    {\delta_A^2\,\delta_B^2}
    \,F(\omega_L;R) \right\}
    \, \hat{s}^{(B)}_Z \,\hat{s}^{(A)}_Z\,,
\end{eqnarray}
\end{widetext}
with \mbox{$\delta_{A/B}=\hbar\omega_L - E_X^{A/B}$} the detuning
for exciton $A/B$ and \mbox{$\delta_{C}=\hbar\,\omega_L - (\hbar\,
c\,q)/n$} the detuning for photons, $c$ the speed of light and $n$
the index of refraction, \mbox{$\textbf{R} =
\textbf{R}_B-\textbf{R}_A$} the distance between donors, and
\begin{eqnarray*} F(\omega_L;R) &=& (2\pi\hbar)^2\,\int
\,d^2\textbf{k}\,
    \frac{\omega_k^2+\left(q\,c/n\right)^2}
    {\omega_k(\omega_L - \omega_k)}\,
    \cos(\textbf{k}\,\textbf{R})\,.
\end{eqnarray*}
The quantity within curly brackets in Eq.~\ref{Eq_Heff} is the
so-called \textit{effective} coupling constant $J_{eff}$.

\subsection{Results}
\label{Sec_2Dres}

\begin{figure}[h]
  \centerline{\includegraphics[scale=1]{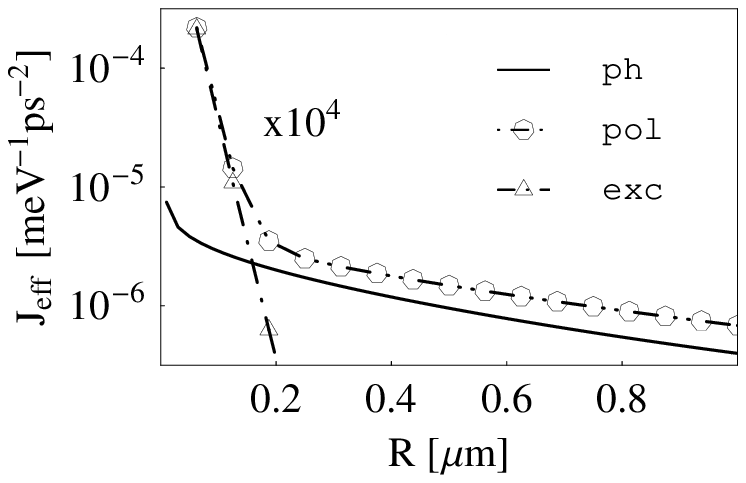}}
  \caption{Effective donor-donor coupling constant $J_{eff}(R)$
  for the interaction mediated by excitons (triangle),
  polaritons (circle) and photons (solid) in a 2D cavity. The parameters
  have been chosen in order to make the correspondence among the three
  cases the most compatible, and to represent a system of Si donors in
  InAs/GaAs QDs: \mbox{$\hbar = 0.625\, meV\, ps$},
  \mbox{$q = 0.021\, nm^{-1}$},
  \mbox{$c = 3.04\,10^5\, nm\, ps^{-1}$},
  \mbox{$\delta_C = -0.66\, meV$}, \mbox{${\cal{V}} = 0.9\, meV$},
  \mbox{$\delta_A = \delta_B = -0.1\,  meV$}, \mbox{$g_A =g_B = 0.3\,
  meV$}, \mbox{$n = 3$}, \mbox{$J_A = J_B = 1\,meV^{-1}ps^{-2}$} and \mbox{$s = 10\,nm$}.
  (For the sake of comparison, the
  polaritons and excitons curves are plotted a factor $10^{-4}$ of their
  actual value)  \label{fig_logJ}}
\end{figure}

For a comprehensive understanding of the results, I first recall
previous similar research works. Theoretical studies have
demonstrated that impurities embedded in a semiconductor host either
in bulk or inside $2$D cavities can be made to interact using
virtual excitons or polaritons+excitons as intermediate particles,
respectively.\cite{piermarocchi_02, piermarocchi_04, quinteiro_06}
These reports show that the spin-spin interaction mediated by
excitons in bulk is Heisenberg, and the interaction mediated by
polaritons+excitons in $2$D cavities is anisotropic (with the
polaritons and excitons providing Ising and transverse terms
respectively). In all cases the coupling strength $J_{eff}$ falls
off exponentially as a function of the separation $R$ between
impurities, having a different range in each case. This exponential
decay is a consequence of virtual (off-resonance) excitations.

Eq.~\ref{Eq_Heff} shows that the donor-donor interaction mediated by
photons is of the Ising type, which can be explained as follows. The
essential ingredient for the correlation between different donors
is, clear enough, the existence of a cavity photon that travels from
one QD to another. The cavity photon acts upon (and is acted on by)
the donors only indirectly, meaning that the photon interacts with
each donor through an \textit{optically active} exciton in the
corresponding QD. No matter how complicated the interaction within
the QD may be (donor's electron-exciton's electron of higher order
or even an interaction that includes a donor's electron-exciton's
hole term), the initial and final states that connect to the cavity
photon must be optically active excitons. This ensures that spin
flips within the QD only occur in pairs of the type
$\hat{s}_+\hat{s}_-$ or $\hat{s}_-\,\hat{s}_+$. According to the
algebra of $SU(2)$, a transformation that is a product of those
pairs is of the form $a+b\,\hat{s}_Z$. Therefore, I conclude that
the restriction of having an interaction mediated only by photons
causes the donor-donor coupling to be Ising-like (to all order).
This is in agreement with the polariton case in $2$D cavities; it is
in contrast to the situation of excitons in bulk or
polaritons+excitons in a $2$D cavity, where the dark excitons also
propagate the interaction and so, do not force the spin flip to come
in pairs of raising and lowering operators for each impurity.

The integration of $F(\omega_L;R)$ is performed numerically after a
cut-off $k_c$ for the in-plane momentum of the photon is prescribed.
The coupling between a cavity photon and a QD exciton
(Eq.~\ref{Eq_HCX}) is taken in the dipole moment approximation. For
this approximation to be valid, the wavelength of the cavity photon
must be larger than the size ``$s$'' of the excitonic wave function
-- roughly equal to the size of a QD -- or, equivalently, the total
momentum $m$ of the photon must be smaller than the inverse of this
size. This imposes a condition on the maximum value of the in-plane
momentum $k$ and yields a cut-off $k_c$ that must satisfy
\mbox{$\sqrt{k_c^2+q^2} \sim 1/s$} -- for $k=0$, the dipole moment
approximation is valid since the width of the cavity is larger than
the size $s$ of the QD.

The effective coupling constant $J_{eff}$ is evaluated using
numerical values for the parameters that represent a system of Si
donors in InAs/GaAs QDs inside a planar microcavity. A plot
comparing the effective coupling $J_{eff}$ for the II mediated by
polaritons, excitons and photons in $2$D cavities is shown in
Fig.~\ref{fig_logJ}. The plot suggests that, for large separation
$R$ between the donors: $i)$ the II decays exponentially -- a
consequence of the off-resonance excitation of the system -- and
$ii)$ the slopes of the polariton and photon coupling constants are
very similar. In order to verify these ideas, we consider the
function $F(\omega_L;R)$ containing the spatial dependence. Its
angular integration yields a Bessel function $J_0(k\,R)$, which
suppresses high momenta $k$ for large $R$. Therefore, the remaining
of the integrand can be expanded in powers of $k/q$, and the
integral can be solved for the lowest order contribution. Then, for
large $R$, $J_{eff}(R) \propto R^{-1/2} exp(-R/R_0)$, with $R_0 =
(2\,M\,\delta_C/\hbar^2)^{-1/2}$ and $M=\hbar\,n\,q/c$ the ``mass''
of the photon. The fitting of $J_{eff}(R)$ by such a Yukawa 2D
function is excellent, and allows to determine the parameter $M$
from the slope; the mass obtained from the fit
$M=\hbar^2/(2\,R_0^2\,\delta_C)$ deviates about $5\%$ from its
theoretical value $M=\hbar\,n\,q/c$. The slope of polaritons can
also be calculated using the Yukawa approximation. The slopes of
both, polaritons and photons, curves agree to $15\%$, and this
mismatch is explained by the slightly different masses of the two
particles. Together with the fact that the polariton and photon
interactions are of the Ising type, I conclude that the long range
coupling of polaritons is of photon nature.

Finally, using an optimistic set of parameters compatible with
experimental studies, I find that the strength of the interaction
mediated by photons is about two orders of magnitude smaller than
that of polaritons. In both cases, the strength of the coupling can
be modified by the detuning $\delta_C$. For the photon mediated
interaction, Eq.~\ref{Eq_Heff} shows that $\delta_A$ and $\delta_B$
play a role too; however, these detunings depend on the exciton
energy of each QD, and they are not simultaneously controllable in
an easy way, and so their manipulation to enhance the photon
mediated II is nowadays doubtful.


\section{The donor-donor interaction mediated by photons
in a $0$D cavity} \label{Sec_0D}

\begin{figure}[h]
  \centerline{\includegraphics[scale=0.35]{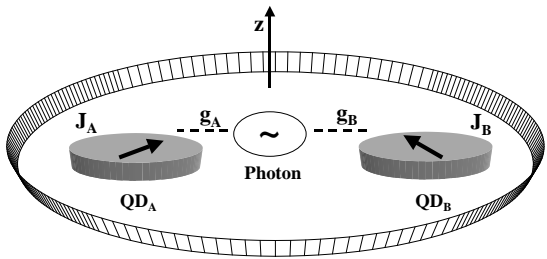}}
  \caption{The 0D cavity model consisting of a 0D-cavity, and two
  QDs each coupled to a donor impurity. \label{fig_0dsys}}
\end{figure}

The system shown in Fig.~\ref{fig_0dsys} consists of two QDs each
coupled to a donor, all embedded in a 0D cavity supporting one
photon mode. The axis normal to the surface of the cavity is named
$z$ and is the quantization axis for the excitons in the QDs for
which the heavy hole (\textit{hh}) and light hole levels split; I
retain the \textit{hh} level only.
I seek to determine the correlation between the spins, and use this
information to infer the indirect interaction in thermal
equilibrium. \cite{Chiappe_05, Chiappe_06} To this end the exact
eigenvalues and eigenvectors of the complete Hamiltonian
\begin{eqnarray}
\label{Eq_H0D}
    H &=& H_0 +
    \sum_{\sigma\,p} g^{(p)} c_\sigma^\dag b_{\sigma}^{(p)} +
    \nonumber \\
    &&
    \sum_{\alpha\alpha^\prime\,p} J^{(p)} s_{\alpha\alpha^\prime}
    \cdot s^{(p)} b_{\alpha\beta}^{(p)\dag}
    b_{\alpha^\prime\beta}^{(p)} + h.c.\,,
\end{eqnarray}
(all orders in $g$ and $J$) were determined by numerical
diagonalization, in the subspace of one particle (photon/$A$
exciton/$B$ exciton)\footnote{The Hamiltonian of the system commutes
with the total number of particles operator, and thus the problem
can be analyzed in a particular subspace of fixed $N$. }. In
Eq.~\ref{Eq_H0D} $c$ and $b$ stand for the operators for cavity
photons and excitons, respectively. The bare energy $H_0$ contains
the exciton energy $E_{X}^p$ of QD $p=A/B$, and the energy $E_c$ of
the cavity photon. Note the similarities between this model and the
$2$D cavity presented in the previous section.

The eigenvectors and energies were then used to calculate the
correlation in a thermal distribution,
\begin{equation*}
    <M_{Aj}M_{Bj}>= \frac{1}{Z} \sum_i e^{-\beta E_i}
        \langle n_i| M_{Aj}M_{Bj} |n_i\rangle  ~,
\end{equation*}
where $n_i$ stands for the eigenvectors of the complete Hamiltonian
of the system, $Z$ is the partition function, and $M_{Dj}$ is the
spin operator in the $j$-direction corresponding to the donor inside
QD $D$. This correlation allows the calculation of the effective
coupling between donors.

\subsection{Results}

First I discuss the validity and scope of this $0$D cavity model.
Consider a typical structure, a micropilar of height \mbox{$h=0.1\,
\mu m$} and diameter \mbox{$\phi=2\, \mu m$}. In this case, the
energy levels of cavity photons are separated by about $\Delta
E\simeq 10 \,meV$, difference much larger than any energy of the
system ($g,\, J<1\, meV$). Therefore, the system presents only one
cavity mode, and the model is applicable to a realistic situation.
In addition, the maximum spatial separation between donors, given by
the size of the micropilar, is at least the separation considered in
the case of a $2$D cavity, so the values of the effective coupling
constant for both models may be compared. Finally, it is worth
noticing that a thorough study of strong and weak coupling between
the QD exciton and the cavity photon would require the inclusion of
decoherence. The model I consider disregards decoherence and so will
only represent the situation of strong coupling ($g$ larger than any
decay constant). In view of the available experimental
data\cite{Ralthmaler_04}, systems of the sort I study here are in
strong coupling regime for values of $g$ in the order of tens of
$\mu eV$ or larger.
\begin{figure}[h]
  \centerline{\includegraphics[scale=1.1]{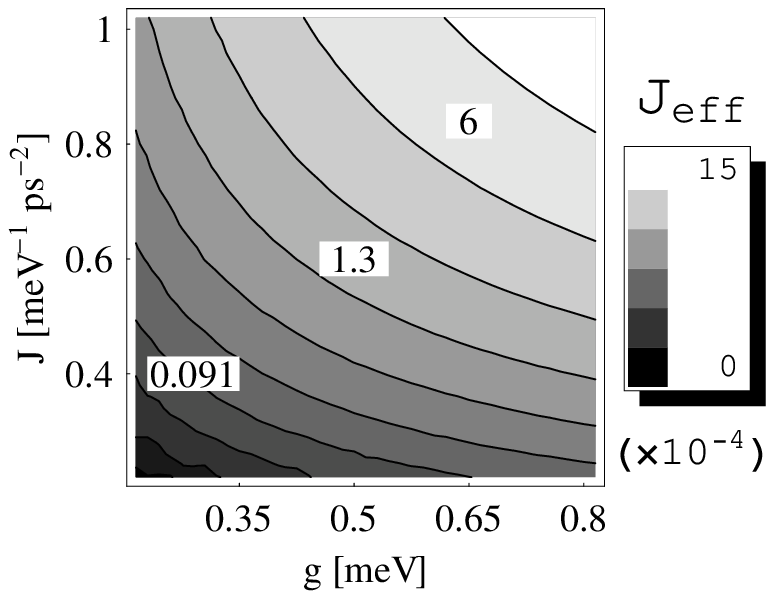}}
  \caption{Contour plot of the Ising effective coupling constant
  $J_{eff}[meV^{-1}ps^{-2}]$ ($\times 10^{-4}$) as a function of couplings
  $J$ and $g$, for a detuning \mbox{$\delta =
  E_{C} - E_{X}^A= 0.5\,meV$} with \mbox{$E_{X}^A=E_{X}^B$},
  \mbox{$g=  g_A=g_B$}, \mbox{$J=J_A=J_B$} and  \mbox{$T=5\,K$}.  \label{Fig_PhaseD}}
\end{figure}

Numerical calculations performed for all admissible values of the
physical constants~\cite{Ralthmaler_04, Chiappe_05, Gywat06} reveal
that the in-plane or $xy$ correlation is zero, consistent with the
results of Sect.~\ref{Sec_2Dres}. Thus I am led to assume an Ising
donor-donor effective Hamiltonian. The numerical data for the normal
or $z$ correlation can be used to infer the value of the effective
coupling constant. A model of two $1/2$ spins interacting through an
Ising Hamiltonian $H=J_{eff}\, s_z^A\,s_z^B$ yields the correlation
\mbox{$<M_{Aj}\,M_{Bj}> = \hbar^2/4 \,\tanh[\hbar^2\,
J_{eff}/(4\,k\,T)]$} in thermal equilibrium. By direct inversion of
this relation or by fitting the data for different temperatures, the
value of the effective coupling constant is computed.

Fig.~\ref{Fig_PhaseD} shows a contour plot of $J_{eff}$ as a
function of the exciton-donor coupling $J$ and photon-exciton
coupling $g$ for an off-resonance situation, at a typical
temperature at which experiments are done. The plot shows, as
expected, that the effective coupling between donors becomes
stronger as the coupling $g$ and $J$ is increased. In addition, I
find that the change of $J_{eff}$ with temperature is very
pronounced, for example at \mbox{$T=0.1K$} the coupling becomes
\mbox{$J_{eff}=2\times10^{-1}\,meV^{-1}\,ps^{-2}$} for
\mbox{$g=0.8\,meV$}, \mbox{$J=1\,meV^{-1}ps^{-2}$}.

Even though the transition from weak to strong coupling can not be
addressed here, it is worth considering what occurs for different
values of the exciton-photon detuning $\delta$. This parameter
causes the excitation in the system to be a photon, an exciton or an
admixture of them: the polariton. To determine the nature of the
excitation, we look at the exact eigenvectors resulting from the
diagonalization of Eq.~\ref{Eq_H0D}. We only need to consider a
small number of them, because at low temperature, the correlation
(and so the $J_{eff}$) is dominated by a small number of low-energy
exact eigenvectors. Every exact eigenvectors is a linear
superposition of vectors of the uncoupled problem $H_0$; the way
these uncoupled vectors mix to form the dominant exact states tells
us the approximate nature of the excitation. To exemplify, take a
set of parameters \{\mbox{$\delta =E_{C} - E_{X}^A = -5\,meV$},
\mbox{$E_{X}^A=E_{X}^B$}, \mbox{$g=0.4\,meV$},
\mbox{$J=0.5\,meV^{-1}ps^{-2}$} and  \mbox{$T=0.1\,K$}\}: the energy
spectrum consists of a set of eight low energy levels well separated
from the rest of the upper levels; the corresponding eight states
dominate the correlation. Every state of this subspace is of the
form $\alpha\ket{C}+...$, with $\alpha\simeq 0.99$ and $\ket{C}$ a
state with a cavity photon (the spin state of the donors is
different for each of the eight eigenvectors).
\begin{figure}[h]
  \centerline{\includegraphics[scale=0.55]{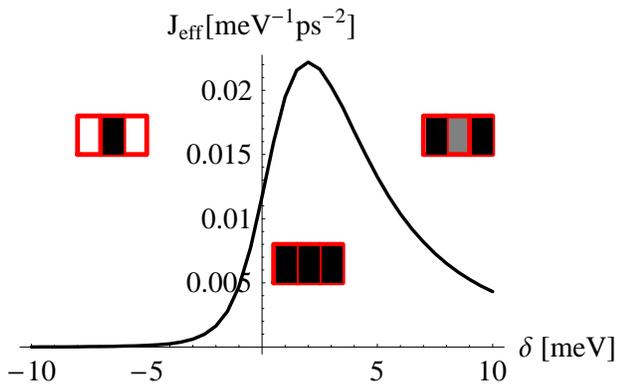}}
  \caption{(color online) $J_{eff}$ vs. detuning \mbox{$\delta =
  E_{C} - E_{X}^A$} for \mbox{$J_A=J_B=0.5\,meV^{-1}ps^{-2}$},
  \mbox{$g_A=g_B=0.8\,meV$}, \mbox{$T=1\,K$} and \mbox{$E_{X}^A=E_{X}^B$}.
  As represented by the three-blocks rectangles, the exact
  eigenstates of the lowest energy subspace are predominantly:
  photon-like for $\delta < 0$, polariton-like for $\delta \simeq 0$,
  and exciton-like for $\delta > 0$.   \label{Fig_JvsDelt}}
\end{figure}
For the probability amplitude, a simple pictorial representation
consisting of a three-blocks rectangle may be used: the left, right
and middle parts of it represent the left QD exciton, the right QD
exciton and the cavity photon, respectively. A gray scale is used,
where the largest probability for the corresponding particle in the
exact eigenstate appears as the darkest block. For the previous
example: $\square \blacksquare \square$. Fig.~\ref{Fig_JvsDelt}
shows the effective coupling constant $J_{eff}$ as a function of the
detuning for \mbox{$E_{X}^A=E_{X}^B$}. For other choices of the
energies \{$E_{C}, E_{X}^A, E_{X}^B$\} one finds  different
compositions of excitations, e.g. \{$E_{C} \simeq E_{X}^B,
E_{X}^A>E_{X}^B$\}: $\square \blacksquare \blacksquare$, a polariton
in the right QD. As can be seen, the largest effective coupling
$J_{eff}$ corresponds to a situation where each of the exact states
belonging to the dominant subspace is an admixture, in equal
proportion, of all the particles of the uncoupled problem $H_0$.

\section{Conclusions}
\label{Sec_Conclusions}

I have studied the donor-donor indirect interaction mediated by
photons in two closely related systems, i.e. a 0D and a 2D cavity. I
found that this spin-spin effective interaction is of the Ising type
and no transverse term is present. The reason for this being that
the donors can interact only through the exchange of a cavity
photon. Photons can only be produced by optically active excitons in
each QD; thus, the initial and final states of each QD are always
optically active excitons. Any spin flip that may occur within the
QD must be compensated by an opposite flip (of the electron or even
the hole if an electron-hole exchange is included in the
Hamiltonian). Because the spin flips come in pairs of raising and
lowering operators, the $s=1/2$ algebra determines that ultimately
the whole operation within each QD is equivalent to $\hat{s}_z$.
This is analogous to what happens in the case of polaritons in a
$2$D cavity. In contrast, excitons in bulk or excitons+polaritons in
a $2$D cavity do not have this restriction and present a transverse
spin-spin interaction.

For the $2$D model, the dependence of the effective coupling
constant ($J_{eff}$) on the donor-donor separation ($R$) plus the
fact that $J_{eff}$ is of the Ising type make evident that the
long-range behavior of polaritons is due to photons. Nevertheless,
for an optimistic set of parameters, the photon mediated interaction
is much weaker than the polariton mediated interaction. One possible
reason for this difference is that the present photon model involves
more interaction steps than the model with polaritons -- polaritons
interact directly with the donor spins, while photons interact only
through the intermediate exciton degree of freedom.

In the case of the $0$D cavity, the strength of the interaction
becomes larger when the exact eigenstate of the system is a full
admixture of all particles: the excitons in each QD and the photon.
Between the situations of positive and negative detuning, the former
seems more effective to maintain correlation among donors. In
contrast to the case of the 2D cavity, $J_{eff}$ does not depend on
the separation between donors, and remains constant up to the values
of $R$ analyzed in the 2D model. The coupling constant is very
sensitive to changes in temperature and it increases as the
temperature decreases.

Although an accurate comparison between the 0D and 2D models is not
possible (since the effective coupling is driven in different ways
in each case), for typical values of the parameters and temperatures
used in experiments, the $J_{eff}$ of the $0$D cavity appears to be
weaker than that of excitons+polaritons. However, for large
separations $R$ and temperatures $T<1\,K$, the numerical results
indicate that the strength of the 0D cavity may overcome that of the
2D system.

From the point of view of application to quantum information
science, these findings suggest that --among the systems compared
here-- either the $0$D-photon or the $2$D-polariton schemes presents
the most advantageous features: the former for the strength at large
spatial separation $R$, and the latter due to the versatility given
by the anisotropic Heisenberg Hamiltonian. Nevertheless, a deep
analysis is still needed to determine how other factors, such as
decoherence, may condition or limit the applications of this
$0$D-photon proposal.

\section{Acknowledgement}

I would like to thank Professors J. Fern\'{a}ndez-Rossier, C.
Piermarocchi, and P. I. Tamborenea. This research was partially
conducted at and financially supported by Michigan State University,
Universidad de Alicante - Spain, and \mbox{ANPCyT} PICT 03-11609 -
Argentina.


\appendix

\section{Effective Hamiltonian for the 2D cavity}
\label{App_Heff}

After applying a transformation to a rotating frame of frequency
$\omega_L$, one can use time-independent perturbation theory on
$H_{LC}$ in the form of projection operators. To second order in
$H_{LC}$, the correction to the unperturbed Hamiltonian is given by
the level-shift operator $R$~\cite{Cohen, piermarocchi_04}
\begin{equation*}
    PRP={\cal{P}}H_{LC}{\cal{Q}}\frac{1}{z-(H_{0}+H_{CX}+H_{XS})}
    {\cal{Q}}H_{LC}{\cal{P}}\, ,
\end{equation*}
where $\cal{P}$ is the projector onto the subspace of zero cavity
photons and excitons, and ${\cal{Q}}=1-{\cal{P}}$. Then,
\begin{equation*}
    H_{eff,\lambda\lambda^\prime} = \langle \lambda| H_{LC}
    \frac{{\cal{Q}}}{{z-(H_{0}+H_{CX}+H_{XS})}}  H_{LC}|\lambda^\prime\rangle
\end{equation*}
where $|\lambda\rangle$ designates any one of
$\{$\mbox{$|\uparrow\uparrow\rangle$},
\mbox{$|\uparrow\downarrow\rangle$},
\mbox{$|\downarrow\uparrow\rangle$},
\mbox{$|\downarrow\downarrow\rangle$}$\}$ vectors of the spin of
both donors, with zero cavity photons and zero excitons. The
light-cavity Hamiltonian acting on $|\lambda\rangle$ yields
\begin{equation*}
    H_{LC}|\lambda\rangle=\hbar \sqrt{{\cal A}} \sum_{\sigma} {\cal{V}}_\sigma
    |\lambda c_0\rangle \, ,
\end{equation*}
with $|\lambda c_0\rangle$ a vector with one cavity photon with
polarization $\sigma$. Then,
\begin{eqnarray*}
    H_{eff,\lambda\lambda^\prime} &=& \hbar^2 {\cal A} \sum_{\sigma}
    |{\cal{V}}_\sigma|^2   \times \\
    &&\langle \lambda c_0 |\frac{1}{{(z-H_{0})-(H_{CX}+H_{XS})}}
    |\lambda^\prime c_0 \rangle\,,
\end{eqnarray*}
with $[G^0_k(\omega_L)]^{-1}=z-H_{0}$. The cavity photon in both
sides of the matrix element are the same, for I seek to study a
coherent process.

I want to consider the case of one virtual particle, either a cavity
photon or an exciton in one or the other QD. That gives three
possible particles. I separate the space of one virtual excitation
into three subspaces $\{ \Im_C, \Im_{A}, \Im_{B}\}$, and obtain
\begin{eqnarray*}
H_I \!\!&=&
  \left(
  \small
  \begin{array}{ccc}
    H_{XS}^A & H_{CX}^A & 0 \\
    H_{CX}^A & 0         & H_{CX}^B \\
    0         & H_{CX}^B & H_{XS}^B \\
  \end{array}
  \right)\\
G^0 \!\!&=&\!\!
  \left(
  \small
  \begin{array}{ccc}
    G^{0A}_{X} & 0       & 0 \\
    0          & G^0_{C} & 0 \\
    0          & 0       & G^{0B}_{X} \\
  \end{array}
  \right)
\end{eqnarray*}
The bare Green's functions are
\begin{eqnarray*}
    G^{0A/B}_X &=& \frac{1}{\hbar\omega_L - E_X^{A/B}} \nonumber \\
    G^{0}_C &=& \frac{1}{\hbar\omega_L - \frac{\hbar c}{n}\sqrt{k^2+q^2}}
\end{eqnarray*}
The first relevant term, in the expansion of the Green's function,
that will give rise to a process that correlates both spins belongs
to the $G^{(2+4)}=G^0 (H_IG^0)^3 (H_IG^0)^3$. The process is: $1)$
creation of an exciton in QD $A$, $2)$ exciton interacting with the
spin, $3)$ annihilation of the exciton in QD $A$ with the creation
of the cavity photon, $4)$ creation of an exciton in QD $B$, $5)$
interaction between exciton and spin, $6)$ annihilation of the
exciton in QD $B$, and creation of the cavity photon. I obtain
\begin{eqnarray*}
    H_{eff,\lambda\lambda^\prime} &=& \hbar^2 {\cal A}
    |G^0_{C\,k=0}|^2 \times \\
    && \sum_{\sigma}|{\cal{V}}_\sigma|^2
    \langle \lambda c_0 |\,\Sigma_{BA}+\Sigma_{AB}\, |\lambda^\prime c_0
    \rangle\,,
\end{eqnarray*}
with the matrix element written in explicit form
\begin{eqnarray*}\label{Eq_Gtilde}
    \langle \lambda c_0 |\,\Sigma_{BA}\,
    |\lambda^\prime c_0 \rangle &=& \langle \lambda c_0 |(H_{CX}^B G^{0B}_{X}
    H_{XS}^B G^{0B}_{X} H_{CX}^B) \times \nonumber \\ && G^{0}_{C} (H_{CX}^A G^{0A}_{X}
    H_{XS}^A G^{0A}_{X} H_{CX}^A) |\lambda^\prime c_0 \rangle \,
\end{eqnarray*}
which is ${\cal{O}}(J^2)$ and ${\cal{O}}(g^4)$. Straight forward,
but lengthy, operator algebra leads to $ H_{eff} = J_{eff} \,
\hat{s}^{(B)}_Z \,\hat{s}^{(A)}_Z$ as presented in the main part of
this article.


\end{document}